\documentclass[footinbib,aps,prb,twocolumn,superscriptaddress,groupedaddress,a4paper]{revtex4} % Uncomment for review/submission version
\usepackage{graphicx}
\usepackage{dcolumn}
\usepackage{bm}
\usepackage{amssymb}
\usepackage{amsmath}
\usepackage{subfigure}
\usepackage{color}
\usepackage{lscape}
\begin{document}

\widetext
\title{Comparison of first principles and semi-empirical models of the\\structural and electronic properties of Ge$_{1-x}$Sn$_{x}$ alloys}

%%%%%%%%%%%%%%%%%%%%%%%%%%%%%%%%%%%%%%%%
%%%% Authors, affiliations and date %%%%
%%%%%%%%%%%%%%%%%%%%%%%%%%%%%%%%%%%%%%%%

\author{Edmond J.~O'Halloran}
\affiliation{Tyndall National Institute, Lee Maltings, Dyke Parade, Cork T12 R5CP, Ireland}
\affiliation{School of Chemistry, University College Cork, Cork T12 YN60, Ireland}

\author{Christopher A.~Broderick}
\email{c.broderick@umail.ucc.ie} % Email address must be kept before affiliation(s)
\affiliation{Tyndall National Institute, Lee Maltings, Dyke Parade, Cork T12 R5CP, Ireland}
\affiliation{Department of Physics, University College Cork, Cork T12 YN60, Ireland}

\author{Daniel S.~P.~Tanner}
\affiliation{Tyndall National Institute, Lee Maltings, Dyke Parade, Cork T12 R5CP, Ireland}

\author{Stefan Schulz}
\affiliation{Tyndall National Institute, Lee Maltings, Dyke Parade, Cork T12 R5CP, Ireland}

\author{Eoin P.~O'Reilly}
\affiliation{Tyndall National Institute, Lee Maltings, Dyke Parade, Cork T12 R5CP, Ireland}
\affiliation{Department of Physics, University College Cork, Cork T12 YN60, Ireland}

\date{\today}

%%%%%%%%%%%%%%%%%%%%%%%%%%%%%%%%%%%%%%%%%%%%%
%%%% Abstract, keywords and PACS numbers %%%%
%%%%%%%%%%%%%%%%%%%%%%%%%%%%%%%%%%%%%%%%%%%%%

\begin{abstract}

We present and compare three distinct atomistic models -- based on first principles and semi-empirical approaches -- of the structural and electronic properties of Ge$_{1-x}$Sn$_{x}$ alloys. Density functional theory calculations incorporating Heyd-Scuseria-Ernzerhof (HSE) and modified Becke-Johnson (mBJ) exchange-correlation functionals are used to perform structural relaxation and electronic structure calculations for a series of Ge$_{1-x}$Sn$_{x}$ alloy supercells. Based on HSE calculations, a semi-empirical valence force field (VFF) potential and $sp^{3}s^{\ast}$ tight-binding (TB) Hamiltonian are parametrised. Comparing the HSE, mBJ and TB models, and using the HSE results as a benchmark, we demonstrate that: (i) mBJ calculations provide an accurate first principles description of the electronic structure at reduced computational cost, (ii) the VFF potential is sufficiently accurate to circumvent the requirement to perform first principles structural relaxation, and (iii) TB calculations provide a good quantitative description of the alloy electronic structure in the vicinity of the band edges. Our results also emphasise the importance of Sn-induced band mixing in determining the nature of the conduction band structure of Ge$_{1-x}$Sn$_{x}$ alloys. The theoretical models and benchmark calculations we present inform and enable predictive, computationally efficient and scalable atomistic calculations for disordered alloys and nanostructures. This provides a suitable platform to underpin further theoretical investigations of the properties of this emerging semiconductor alloy.

\end{abstract}

% \keywords{0.0X}
% \pacs{0.0X}

\maketitle

%%%%%%%%%%%%%%%%%%%%%%
%%%% Introduction %%%%
%%%%%%%%%%%%%%%%%%%%%%

\section{Introduction}
\label{sec:introduction}

The group-IV semiconductors silicon (Si) and germanium (Ge) underpin modern electronics but are unsuitable for applications in active photonic devices due to their indirect band gaps, which make them intrinsically inefficient emitters and absorbers of light. As Moore's Law approaches its end there is a strong impetus to develop novel hybrid optoelectronic architectures, to deliver continued enhancement in processing speeds and capacity. \cite{Zhou_LSAA_2015,Winzer_JLT_2017} Significant research effort has been devoted over the past two decades to Si photonics: the development of photonic devices which are compatible with established complementary metal-oxide semiconductor (CMOS) processing infrastructure, to enable monolithic integration with microelectronics and deliver step-changes in device performance and capabilities. \cite{Thomson_JO_2016} While much progress has been made in the development of passive photonic components, the development of Si photonics is limited by the unavailability of CMOS-compatible direct-gap materials suitable for applications in semiconductor lasers and light-emitting diodes. \cite{Zhou_LSAA_2015,Geiger_FM_2015}

Recently, there has been growing interest in alloy group-IV materials to achieve CMOS-compatible direct-gap semiconductors. Due to the fact that its (fundamental) indirect $\Gamma_{8v}$-L$_{6c}$ band gap is only $\approx 145$ meV smaller than its direct band gap, engineering the conduction band (CB) structure of Ge has attracted significant attention. \cite{Geiger_FM_2015,Saito_SST_2016,Reboud_PCGC_2017} The most prominent current approach is alloying of Ge with tin (Sn). \cite{Kouvetakis_ARMR_2006,Soref_PTRSA_2014,Zaima_STAM_2015} Originally proposed over two decades ago, Ge$_{1-x}$Sn$_{x}$ alloys have recently seen a surge of interest due to initial demonstrations of optically \cite{Wirths_NP_2015} and electrically \cite{Margetis_APL_2018} pumped lasing. In addition to applications in CMOS-compatible light-emitting devices, research interest in (Si)Ge$_{1-x}$Sn$_{x}$ alloys has also been driven by potential applications in tunneling field-effect transistors \cite{Yang_PIEDM_2012,Wirths_APL_2013,Santa_APL_2014} and multi-junction solar cells. \cite{Fang_JACS_2008,Roucka_IEEEJPV_2016,Timo_AIPCP_2018}

Comparison by Moontragoon et al.~\cite{Moontragoon_SST_2007} between empirical pseudopotential calculations for Ge$_{1-x}$Sn$_{x}$ carried out using the virtual crystal approximation (VCA), and a direct atomistic (alloy supercell) approach, highlighted significant differences in key properties including band gap bowing. This suggested the potential importance of band mixing and alloy disorder effects in Ge$_{1-x}$Sn$_{x}$ alloys. More recent analyses have provided further evidence of the strong role played by band mixing effects in the determining the nature of the Ge$_{1-x}$Sn$_{x}$ electronic structure. Eckhardt et al.~\cite{Eckhardt_PRB_2014} noted the emergence in density functional theory (DFT) supercell calculations of a strongly hybridised alloy CB edge state, consisting of an admixture of the $\Gamma_{7c}$ and L$_{6c}$ CB edge states of Ge. The unfolded supercell band structures presented by Polak et al.~\cite{Polak_JPDAP_2017} also displayed clear indications of band mixing. More recently, hydrostatic pressure measurements -- supported by DFT calculations -- also indicated the presence of strong Sn-induced $\Gamma_{7c}$-L$_{6c}$ mixing, \cite{Eales_PW_2017,Eales_submitted_2019} suggesting (i) that the band gap pressure coefficient $\frac{ \textrm{d} E_{g} }{ \textrm{d} P }$ in Ge$_{1-x}$Sn$_{x}$ is intermediate between that of the indirect $\Gamma_{8v}$-L$_{6c}$ and direct band gaps of Ge up to $x = 10$\%, and (ii) that $\frac{ \textrm{d} E_{g} }{ \textrm{d} P }$ evolves continuously with increasing $x$, from that of the indirect Ge $\Gamma_{8v}$-L$_{6c}$ band gap at $x = 0$ towards that of the direct Ge band gap at $x = 10$\%.

While not yet conclusive -- for example, the pressure-dependent measurements of Dybala et al.~\cite{Dybala_JAP_2016} suggested little Sn composition dependence of $\frac{ \textrm{d} E_{g} }{ \textrm{d} P }$ -- there is a growing body of evidence to suggest that Sn-induced hybridisation of states lying close in energy to the Ge CB edge is likely to be of significant importance in determining the nature of the electronic structure evolution in Ge$_{1-x}$Sn$_{x}$. For example, the presence of band mixing indicates that a direct band gap evolves continuously with increasing Sn composition,\cite{Eales_PW_2017,Eales_submitted_2019,Broderick_IEEENano_2018,Schulz_NUSOD_2018,Broderick_NUSOD_2019} in stark contrast to the general assumption in the literature of an abrupt transition to a direct band gap at some critical Sn composition $x$. The presence of such band mixing effects will have significant implications for technologically relevant material properties, but such effects have to date received little attention in the literature. Indeed, despite early demonstrations\cite{He_PRL_1997,Yin_PRB_2008} of the breakdown of the VCA, there has been a proliferation of VCA-based approaches to interpret the electronic and optical properties of Ge$_{1-x}$Sn$_{x}$.

Given the strong interest in (Si)Ge$_{1-x}$Sn$_{x}$ for practical applications, a suitable theoretical platform is required not only to facilitate quantification of the importance of alloying effects in Ge$_{1-x}$Sn$_{x}$ alloys, but also to underpin direct atomistic calculations of the properties of realistically-sized Ge$_{1-x}$Sn$_{x}$-based nanostructures. Such analysis will not only quantify the evolution of the alloy electronic structure, but will inform quantitative calculations of technologically relevant material properties and, ultimately, the development of predictive theoretical models which can be utilised in the design and optimisation of photonic, electronic and photovoltaic devices based on Ge$_{1-x}$Sn$_{x}$ and related alloys.

Our aim in this paper is to establish and benchmark such a platform. We present and compare three distinct first principles and semi-empirical atomistic models to compute the structural and electronic properties of Ge$_{1-x}$Sn$_{x}$ alloys. The first, DFT-based model uses a highly accurate but computationally expensive hybrid functional scheme, employing the Heyd-Scuseria-Ernzerhof (HSE) exchange-correlation (XC) functional \cite{Heyd_JCP_2003,Heyd_JCP_2004} modified for solids (HSEsol). \cite{Schimka_JCP_2011} The second, DFT-based model uses the local density approximation (LDA) for structural relaxation, with electronic structure calculations then performed using the modified Becke-Johnson (mBJ) XC functional. \cite{Tran_PRL_2009} This second model is similar to that employed in recent investigations of Ge$_{1-x}$Sn$_{x}$ alloys. \cite{Eckhardt_PRB_2014,Polak_JPDAP_2017} The third, semi-empirical model combines structural relaxation based on a valence force field (VFF) potential with electronic structure calculations based on a nearest-neighbour $sp^{3}s^{\ast}$ tight-binding (TB) Hamiltonian with both the VFF potential and TB Hamiltonian parametrised directly using HSEsol-calculated structural properties, band structures and band edge deformation potentials.

Using the HSEsol calculations as a benchmark we quantify the accuracy of structural relaxations and electronic structure calculations carried out for a series of Ge$_{1-x}$Sn$_{x}$ alloy supercells using the LDA + mBJ and VFF + TB models. We find that the LDA + mBJ model offers first principles calculations of the alloy properties which are in good quantitative agreement with, and come at significantly reduced computational cost compared to HSEsol calculations. We also demonstrate that the semi-empirical VFF + TB model offers an accurate and computationally inexpensive approach to calculate the structural and electronic properties, describing well the alloy electronic structure close in energy to the CB and valence band (VB) edges -- i.e.~in the regions of the band structure of interest to calculate technologically relevant material properties such as optical transition strengths, carrier mobilities and band-to-band tunneling rates.

We find that the VFF structural relaxations are in excellent quantitative agreement with full HSEsol calculations, suggesting that VFF relaxation can circumvent the requirement to carry out first principles structural relaxations -- i.e.~the relaxed positions obtained from a VFF structural relaxation can be reliably used as input to first principles electronic structure calculations, offering significant reductions in computational cost. For equivalent system sizes we typically find that the computational cost associated with mBJ calculations is reduced by approximately an order of magnitude compared to equivalent HSEsol calculations, while semi-empirical VFF + TB calculations on the same supercells come at negligible computational cost. While mBJ calculations allow access to larger system sizes in first principles calculations due to their reduced computational cost, they are nonetheless limited to systems containing $\lesssim 10^{3}$ atoms. Conversely, the semi-empirical VFF + TB model we establish is highly scalable, and can readily be applied to systems containing $\sim 10^{6}$ atoms. This provides significant scope for future investigations, to underpin atomistic calculations of the properties of disordered Ge$_{1-x}$Sn$_{x}$ alloys and realistically-sized nanostructures.

The remainder of this paper is organised as follows. In Sec.~\ref{sec:theoretical_models} we describe the theoretical approaches we have established to calculate the structural and electronic properties of Ge$_{1-x}$Sn$_{x}$ alloys. Sec.~\ref{sec:models_dft} describes the HSEsol and mBJ DFT calculations, while the semi-empirical VFF potential and TB Hamiltonian are described in Sec.~\ref{sec:models_semi_empirical}. The results of our calculations are presented in Sec.~\ref{sec:results}, beginning in Sec.~\ref{sec:results_relaxation} with the relaxation of atomic positions in Ge$_{1-x}$Sn$_{x}$ alloy supercells, before describing their electronic structure in Sec.~\ref{sec:results_band_structure}. Finally, in Sec.~\ref{sec:conclusions} we summarise and conclude.

%%%%%%%%%%%%%%%%%%%%%%%%%%%%
%%%% Theoretical models %%%%
%%%%%%%%%%%%%%%%%%%%%%%%%%%%

\section{Theoretical models}
\label{sec:theoretical_models}

In this section we describe the theoretical models we have established to analyse Ge$_{1-x}$Sn$_{x}$ alloys. We begin in Sec.~\ref{sec:models_dft} by describing the implementation of the HSEsol and mBJ DFT calculations. The HSEsol calculations are as described in Ref.~\onlinecite{Eales_submitted_2019}: we recapitulate their details for completeness. In Sec.~\ref{sec:models_semi_empirical} we describe a semi-empirical model which respectively employs a VFF potential and TB Hamiltonian to perform structural relaxations and electronic structure calculations. In Sec.~\ref{sec:model_supercells} we describe the choice of supercells used to perform our benchmark calculations.

%%%%%%%%%%%%%%%%%%%%%%%%%%%%%%%%%%%%%%%%%%%%%%%%%%%%%%%%%%%%%%%%%%%%%%%%%%%%%%%%%%
%%%% First principles: Heyd-Scuseria-Ernzerhof and modified Becke-Johnson DFT %%%%
%%%%%%%%%%%%%%%%%%%%%%%%%%%%%%%%%%%%%%%%%%%%%%%%%%%%%%%%%%%%%%%%%%%%%%%%%%%%%%%%%%

\subsection{First principles: Heyd-Scuseria-Ernzerhof\\and modified Becke-Johnson DFT}
\label{sec:models_dft}

For the HSE and mBJ DFT calculations we employ pseudopotentials in which the (4$s$)$^{2}$(4$p$)$^{2}$ states of Ge and (5$s$)$^{2}$(5$p$)$^{2}$ orbitals of Sn are explicitly treated as valence states. The semi-core $d$ states of both Ge and Sn are treated as core electron states, since unfreezing these states has been demonstrated to have a negligible impact on the calculated electronic structure. \cite{Eckhardt_PRB_2014} Due to the large relativistic effects associated with Sn, and to a lesser extent with Ge, all calculations explicitly include spin-orbit coupling. All DFT calculations were performed using the projector augmented-wave method, \cite{Blochl_PRB_1994,Kresse_PRB_1999} as implemented in the Vienna Ab-initio Simulation Package (VASP). \cite{Kresse_PRB_1996,Kresse_CMS_1996} For primitive unit cells we utilise a $\Gamma$-centred $6 \times 6 \times 6$ Monkhorst-Pack \textbf{k}-point grid for Brillouin zone integration, which is downsampled appropriately for larger supercells in order to preserve the resolution of the Brillouin zone sampling. Structural relaxation is achieved via free energy minimisation, by allowing the lattice vectors and internal degrees of freedom to relax freely, subject to the additional criterion that the maximum force on any atom in the supercell does not exceed 0.01 eV \AA$^{-1}$.

% HSEsol

For the hybrid functional calculations we employ the HSE XC functional \cite{Heyd_JCP_2003,Heyd_JCP_2004} modified for solids (HSEsol). \cite{Schimka_JCP_2011} A screening parameter $\mu = 0.2$ \AA$^{-1}$ is used for all calculations. We use a high plane wave cut-off energy of 400 eV, to minimise Pulay stress and allow for accurate supercell relaxation. Since Sn incorporation is known to strongly impact states lying close in energy to the Ge CB edge, our priority in establishing the HSEsol calculations was to accurately describe the separation in energy between the $\Gamma_{7c}$ and L$_{6c}$ CB edge states of Ge. We therefore treat the exact exchange mixing $\alpha$ as an adjustable parameter, which is chosen so that the calculated band structure reproduces closely the experimentally measured $\Gamma_{7c}$-L$_{6c}$ splitting. It was found that $\alpha = 0.3$ -- i.e.~30\% exact exchange mixing -- reproduces this splitting accurately. Since we are interested in experimentally relevant Sn compositions $x \lesssim 20$\%, this value is therefore adopted for all HSEsol calculations on Ge$_{1-x}$Sn$_{x}$ alloy supercells.

% mBJ

For the mBJ calculations the same plane wave cut-off energy and \textbf{k}-point grids were employed as for the HSEsol calculations, to allow the two approaches to be compared on an equal basis. In the mBJ calculations we again prioritise the description of the lowest CB in Ge: the parameter $c$ in the mBJ XC functional \cite{Tran_PRL_2009} is treated as adjustable, and chosen to reproduce the experimentally measured $\Gamma_{7c}$-L$_{6c}$ splitting. It was found that this is achieved for $c = 1.2$, which is therefore adopted for all mBJ calculations on Ge$_{1-x}$Sn$_{x}$ alloy supercells.

% Table 1

\begin{table*}[t!]
	\caption{\label{tab:dft_benchmark} Lattice constant $a$, direct band gap $E_{g}$ and VB spin-orbit splitting energy $\Delta_{\protect\scalebox{0.6}{\textrm{SO}}}$ for Ge, $\alpha$-Sn and zb-GeSn, calculated via DFT using the HSEsol (with $\alpha = 0.3$), and LDA (for $a$) or mBJ (with $c = 1.2$, for $E_{g}$ and $\Delta_{\protect\scalebox{0.6}{\textrm{SO}}}$) XC functionals, and compared to low-temperature experimental measurements and previous first principles theoretical calculations. For Ge the fundamental (indirect) L$_{6c}$-$\Gamma_{8v}$ band gap is listed in parentheses.}
	\begin{ruledtabular}
		\begin{tabular}{c|ccc|ccc|ccc}
			            &        & $a$ (\AA) &                          &               & $E_{g}$ (eV)  &                     &        & $\Delta_{\scalebox{0.6}{\textrm{SO}}}$ (eV) &             \\
			Material    & HSEsol & LDA       & Reference                & HSEsol        & mBJ           & Reference           & HSEsol & mBJ                                         & Reference   \\
			\hline
			Ge          & 5.646  & 5.647     & 5.657$^{a}$, 5.648$^{b}$ & 0.908 (0.765) & 0.868 (0.724) & 0.890 (0.744)$^{d}$ & 0.322  & 0.274                                       & 0.296$^{d}$ \\
			$\alpha$-Sn & 6.496  & 6.483     & 6.489$^{a}$              & $-$0.382      & $-$0.401      & $-$0.413$^{e}$      & 0.750  & 0.651                                       & 0.800$^{e}$ \\
			zb-GeSn     & 6.079  & 6.074     & 6.127$^{c}$, 6.032$^{f}$ & 0.040         & 0.007         & 0.085$^{f}$         & 0.460  & 0.424                                       & 0.480$^{f}$ \\
		\end{tabular}
	\end{ruledtabular}
	\begin{flushleft}
	$^{a}$Meas.~average, Ref.~\onlinecite{Landolt_1982_1} \;
	$^{b}$Calc.~average, Ref.~\onlinecite{Landolt_1982_1} \;
	$^{c}$Calc.~average, Refs.~\onlinecite{Brudevoll_PRB_1993,Rucker_PRB_1995,Pandey_APL_1999,Khenata_PB_2003} \;
	$^{d}$Ref.~\onlinecite{Landolt_1982_2} \;
	$^{e}$Ref.~\onlinecite{Groves_JPCS_1970} \;
	$^{f}$Ref.~\onlinecite{Brudevoll_PRB_1993} \;
	\end{flushleft}
\end{table*}

% Comparison to experiment and previous theory

The lattice constants, band gaps and VB spin-orbit splitting energies calculated using the HSEsol and LDA/mBJ XC functionals -- for the constituent elemental materials Ge and $\alpha$-Sn, as well as the fictitious zinc blende IV-IV compound GeSn (zb-GeSn) -- are summarised in Table~\ref{tab:dft_benchmark}, where they are compared to low temperature experimental measurements and previous theoretical calculations. For Ge the HSEsol-calculated direct $\Gamma_{7c}$-$\Gamma_{8v}$ and indirect L$_{6c}$-$\Gamma_{8v}$ band gaps are overestimated by $\approx 20$ meV compared to experiment, but the corresponding $\Gamma_{7c}$-L$_{6c}$ splitting of 143 meV is in excellent agreement with experiment. The corresponding mBJ-calculated direct and indirect band gaps for Ge are underestimated by $\approx 20$ meV compared to experiment, but the corresponding $\Gamma_{7c}$-L$_{6c}$ splitting of 144 meV is again in excellent agreement with experiment. The HSEsol (mBJ) calculated VB spin-orbit splitting overestimates (underestimates) the measured value by 26 meV (22 meV). For $\alpha$-Sn the magnitude of the HSEsol (mBJ) calculated direct band gap underestimates the measured value by 31 meV (12 meV), while the VB spin-orbit splitting is underestimated by 50 meV (149 meV). Our calculated lattice constant, direct band gap and VB spin-orbit splitting energy are in good agreement with previous theoretical calculations. \cite{Brudevoll_PRB_1993} The HSEsol- and mBJ-calculated band structures of Ge, $\alpha$-Sn and zb-GeSn are shown, using solid black and dotted red lines respectively, in Figs.~\ref{fig:band_structure_comparison}(a),~\ref{fig:band_structure_comparison}(b) and~\ref{fig:band_structure_comparison}(c). We note that, compared to the HSEsol XC functional, the mBJ XC functional tends to (i) overestimate the zone-centre effective mass of the lowest energy CB, (ii) underestimate the magnitude of the VB spin-orbit splitting energy, and (iii) underestimate the energies of higher lying zone centre CB states.

%%%%%%%%%%%%%%%%%%%%%%%%%%%%%%%%%%%%%%%%%%%%%%%%%%%%%%%%%%%%%%%
%%%% Semi-empirical: valence force field and tight-binding %%%%
%%%%%%%%%%%%%%%%%%%%%%%%%%%%%%%%%%%%%%%%%%%%%%%%%%%%%%%%%%%%%%%

\subsection{Semi-empirical: valence force field potential\\and $sp^{3}s^{\ast}$ tight-binding Hamiltonian}
\label{sec:models_semi_empirical}

% VFF

% \subsubsection{Valence force field potential}

Our semi-empirical description of the structural and elastic properties of Ge$_{1-x}$Sn$_{x}$ alloys is based on a nearest-neighbour VFF potential. The VFF potential we employ is that originally introduced by Musgrave and Pople, \cite{Musgrave_PRSLA_1962} and later modified by Martin. \cite{Martin_PRB_1970} We consider this potential in its non-polar form, whereby the contribution to the lattice free energy associated with an atom located at site $i$ is given by

\begin{align}
	V_{i} &= \frac{1}{2} \sum_{j} \frac{ k_{r} }{2} \left( r_{ij} - r_{ij}^{(0)} \right)^{2} \nonumber \\
	&+ \sum_{j} \sum_{k > j} \bigg[ \frac{ k_{\theta} }{2} r_{ij}^{(0)} r_{ik}^{(0)} \left( \theta_{ijk} - \theta_{ijk}^{(0)} \right)^{2} \nonumber \\
	&\hspace{1.5cm}+ k_{rr} \left( r_{ij} - r_{ij}^{(0)} \right) \left( r_{ik} - r_{ik}^{(0)} \right) \nonumber \\
	&+ k_{r\theta} \bigg( r_{ij}^{(0)} \left( r_{ij} - r_{ij}^{(0)} \right) + r_{ik}^{(0)} \left( r_{ik} - r_{ik}^{(0)} \right) \bigg) \left( \theta_{ijk} - \theta_{ijk}^{(0)} \right) \bigg] \, ,
	\label{eq:vff_potential}
\end{align}

\noindent
where the indices $j$ and $k$ describe the nearest-neighbour atoms of atom $i$. The unstrained (equilibrium) and relaxed bond lengths between atoms $i$ and $j$ are denoted respectively by $r_{ij}^{(0)}$ and $r_{ij}$; $\theta_{ijk}^{(0)}$ and $\theta_{ijk}$ respectively denote the unstrained and relaxed angles between the adjacent nearest neighbour bonds formed by atoms $i$ and $j$, and $i$ and $k$. The first and second terms in Eq.~\eqref{eq:vff_potential} respectively describe contributions to the lattice free energy associated with pure bond stretching (changes in $r_{ij}$) and pure bond-angle bending (changes in $\theta_{ijk}$), while the third and fourth terms are ``cross terms'' which respectively describe the impact of changes in $r_{ik}$ on $r_{ij}$, and the impact of changes in $\theta_{ijk}$ on both $r_{ij}$ and $r_{ik}$.

By re-casting the lattice free energy expressed in Eq.~\eqref{eq:vff_potential} in terms of macroscopic and internal strains, and inverting the associated derived expressions for the elastic constants $C_{11}$, $C_{12}$ and $C_{44}$, and Kleinman (internal strain) parameter $\zeta$, we have derived analytical expressions describing the force constants $k_{r}$, $k_{\theta}$, $k_{rr}$ and $k_{r\theta}$ explicitly in terms of $C_{11}$, $C_{12}$, $C_{44}$, $\zeta$ and the equilibrium bond length $r^{(0)}$. The derivation of these analytical relationships is outlined in Refs.~\onlinecite{Tanner_thesis_2018} and~\onlinecite{Tanner_VFF_2019}. These analytical relationships circumvent the conventional requirement to perform numerical fitting to determine the force constants, and provide an exact description of the static lattice properties of the constituent materials Ge, $\alpha$-Sn and zb-GeSn in the linear elastic limit. The parameters used in our VFF structural relaxations are provided in Table~\ref{tab:vff_parameters}. Full details of the parametrisation of Eq.~\eqref{eq:vff_potential} based on HSEsol calculations will be provided in Ref.~\onlinecite{Tanner_elastic_2018}. To perform structural relaxations for Ge$_{1-x}$Sn$_{x}$ alloy supercells this VFF potential was implemented using the General Utility Lattice Program (GULP) \cite{Gale_JCSFT_1997,Gale_MS_2003,Gale_ZK_2005} where, as in the DFT calculations described above, the structural relaxation proceeds by minimising the lattice free energy.

% TB

% \subsubsection{Tight-binding Hamiltonian}

Our semi-empirical electronic structure calculations for Ge$_{1-x}$Sn$_{x}$ alloys are based on a nearest-neighbour $sp^{3}s^{\ast}$ TB Hamiltonian, including spin-orbit coupling. The unstrained band structures of Ge, $\alpha$-Sn and zb-GeSn are parametrised based on those calculated using the HSEsol formalism described in Sec.~\ref{sec:models_dft}. As with the HSEsol and mBJ DFT calculations, our priority in generating a TB fit to the Ge band structure is to provide an accurate description of the lowest energy CB, in terms of the energies of the CB edge states at the L, $\Gamma$ and X points in the Brillouin zone. We begin by following the fitting procedure outlined in Ref.~\onlinecite{Vogl_JPCS_1983}, which relates the TB parameters to the energies at the $\Gamma$ and X points, and then adjust the resulting parameters to achieve an exact fit to the energies of the L$_{6c}$, $\Gamma_{7c}$ and $X_{5c}$ CB edge states. For the $\alpha$-Sn band structure we apply the fitting procedure of Ref.~\onlinecite{Vogl_JPCS_1983} without modification, which we find provides a good overall fit to the HSEsol-calculated band structure.

% Table 2

\begin{table}[b!]
	\caption{\label{tab:vff_parameters} Equilibrium bond lengths $r^{(0)}$, and force constants $k_{r}$, $k_{\theta}$, $k_{rr}$ and $k_{r\theta}$, used to implement structural relaxations using the VFF potential of Eq.~\eqref{eq:vff_potential} for Ge$_{1-x}$Sn$_{x}$ alloy supercells. Force constants have been computed analytically based on HSEsol-calculated structural properties for Ge, $\alpha$-Sn and zb-GeSn. \cite{Tanner_VFF_2019,Tanner_elastic_2018} }
	\begin{ruledtabular}
		\begin{tabular}{clccc}
			Parameter     & Unit                     & Ge     & $\alpha$-Sn & zb-GeSn   \\
			\hline
			$r^{(0)}$     & \AA                      & 2.445  &   2.813     & 2.632     \\
			$k_{r}$       & eV \AA$^{-2}$            & 7.0414 &   6.5920    & 7.7970    \\
			$k_{\theta}$  & eV \AA$^{-2}$ rad$^{-2}$ & 0.5104 &   0.2594    & 0.3375    \\
			$k_{rr}$      & eV \AA$^{-2}$            & 0.2416 & $-0.1246$   & $-0.2369$ \\
			$k_{r\theta}$ & eV \AA$^{-2}$ rad$^{-1}$ & 0.3005 &   0.0430    & $-0.0753$ \\
		\end{tabular}
	\end{ruledtabular}
\end{table}

% Figure 1

\begin{figure*}[t!]
	\includegraphics[width=0.99\textwidth]{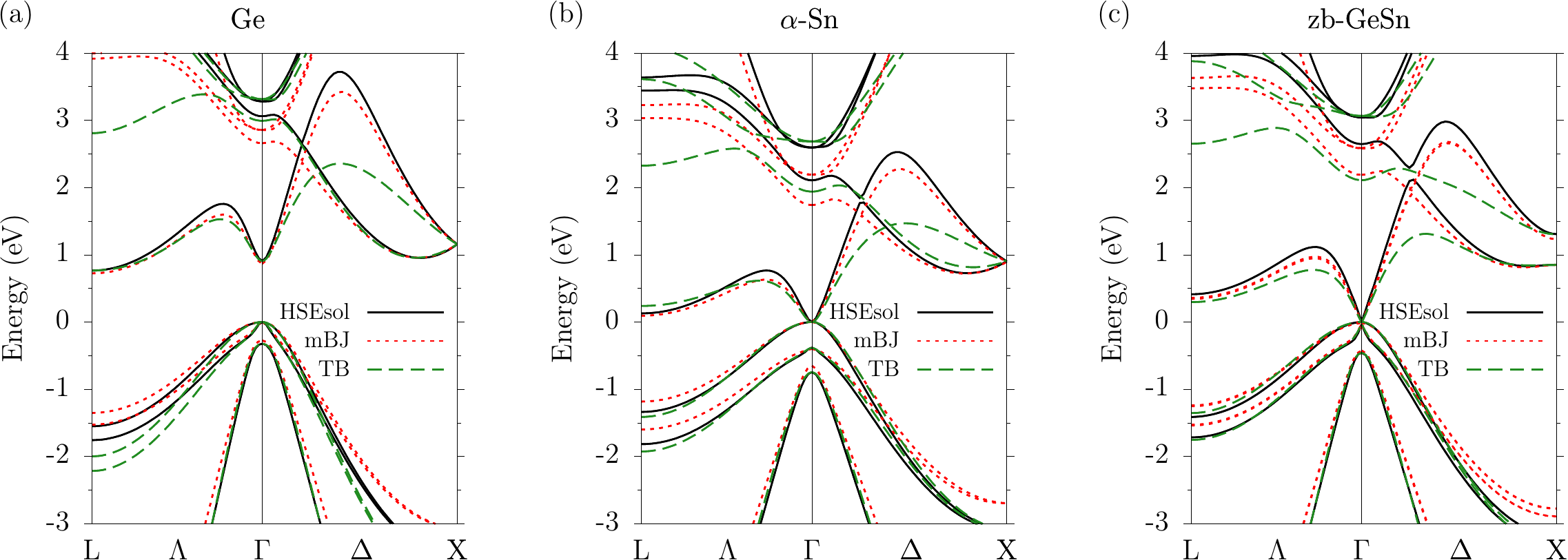}
	\caption{Band structure of (a) Ge, (b) $\alpha$-Sn, and (c) zb-GeSn calculated via DFT, using the HSEsol (solid black lines) and mBJ (dotted red lines) XC functionals, and via a semi-empirical $sp^{3}s^{\ast}$ TB Hamiltonian (dashed green lines). All calculations include spin-orbit coupling. For comparative purposes, the zero of energy has been chosen to lie at the VB edge in all cases.}
	\label{fig:band_structure_comparison}
\end{figure*}

Key to describing the impact of Sn incorporation on the electronic structure of Ge$_{1-x}$Sn$_{x}$ alloys is the TB fit to the band structure of a zb-GeSn primitive unit cell. Such a fitting allows for an accurate description of the interaction between orbitals situated on neighbouring Ge and Sn atoms in alloy supercell calculations. To generate a suitable set of TB parameters for zb-GeSn we again begin with the procedure outlined in Ref.~\onlinecite{Vogl_JPCS_1983}. Following this procedure provides a good overall fit to the HSEsol-calculated zb-GeSn band structure, but we find that -- compared to HSEsol and mBJ calculations -- these parameters tend to underestimate the impact of Sn incorporation in Ge$_{1-x}$Sn$_{x}$ alloy supercell calculations. To rectify this we adjust the differences in the free atomic orbital energies used to determine the Ge and Sn atomic orbital energies in the zb-GeSn TB Hamiltonian. \cite{Vogl_JPCS_1983} In doing so, a more accurate description of the impact of Sn incorporation is obtained, while simultaneously maintaining a good overall fit to the HSEsol-calculated zb-GeSn band structure. The adjusted TB fit for zb-GeSn described above leads to significant overestimation of the effective masses in the lowest CB but, as we will describe in Sec.~\ref{sec:results_band_structure} below, allows for an accurate description of the electronic structure of Ge$_{1-x}$Sn$_{x}$ alloy supercells. The dashed green lines in Figs.~\ref{fig:band_structure_comparison}(a),~\ref{fig:band_structure_comparison}(b) and~\ref{fig:band_structure_comparison}(c) respectively show the band structure of Ge, $\alpha$-Sn and zb-GeSn calculated using the $sp^{3}s^{\ast}$ TB Hamiltonian. The TB fits for Ge and $\alpha$-Sn compare well overall to the HSEsol-calculated band structures, with the quality of the fit reducing for energies $\gtrsim 1$ eV above the CB edge (behaviour typical of an $sp^{3}s^{\ast}$ model \cite{Vogl_JPCS_1983}).

The impact of strain is incorporated in the TB Hamiltonian by describing the dependence of the inter-atomic interaction matrix elements on changes in nearest-neighbour bond lengths and angles using, respectively, the generalised form of Harrison's rule and the Slater-Koster two-centre integrals. \cite{Slater_PR_1954} We include two additional strain-related terms in the TB Hamiltonian to correct the deformation potentials associated with applied tetragonal (biaxial) deformations, which we have re-cast in terms of the local strain at each lattice site to facilitate their incorporation in supercell calculations. Firstly, we include an on-site correction of the $p$ orbital energies using an axial deformation potential $b_{p}$, which provides an accurate description of $\Gamma$-point VB edge axial deformation potential $b$. \cite{Rucker_PSSB_1983} Secondly, we include a correction to the $V_{s^{\ast}p\sigma}$ inter-atomic interaction matrix elements, which accounts for the influence of $d$ orbital interactions in determining the axial deformation potential $\Xi_{u}^{\text{X}}$ associated with the X-point CB edge states. \cite{Munoz_PRB_1993}

For the Ge$_{1-x}$Sn$_{x}$ alloy supercell calculations the atomic orbital energies and inter-atomic interaction matrix elements are computed depending explicitly on the local neighbour environment. The orbital energies at a given lattice site $i$ are determined by averaging over the orbital energies of the materials -- Ge, $\alpha$-Sn or zb-GeSn -- formed by atom $i$ and each of its four nearest neighbours $j$ (including the corresponding VB offsets). We assume a natural VB offset of 1.02 eV between Ge and $\alpha$-Sn, following the first principles calculations of Li et al. \cite{Li_APL_2009} Based on a linear interpolation of this VB offset with respect to the HSEsol-calculated lattice constants, we estimate a VB offset of 0.51 eV between Ge and zb-GeSn. The inter-atomic interaction matrix elements between nearest-neighbour atoms are constructed as outlined above, using the relaxed nearest-neighbour bond lengths and angles. This explicitly local construction of the supercell Hamiltonian allows for direct and accurate incorporation of alloy disorder effects in electronic structure calculations -- via explicitly incorporation of atomic size and chemical differences, in addition to lattice relaxation -- and thus for a quantitative analysis of their impact on the material properties. Full details of the parametrisation of the TB Hamiltonian for Ge$_{1-x}$Sn$_{x}$ and related group-IV alloys will be provided in Ref.~\onlinecite{Broderick_electronic_2018}.

%%%%%%%%%%%%%%%%%%%%%%%%%%%%%%%%%%%%
%%%% Choice of alloy supercells %%%%
%%%%%%%%%%%%%%%%%%%%%%%%%%%%%%%%%%%%

\subsection{Choice of alloy supercells}
\label{sec:model_supercells}

Emerging theoretical \cite{Eckhardt_PRB_2014} and experimental \cite{Eales_PW_2017,Eales_submitted_2019} evidence suggests the possibility of strong Sn-induced hybridisation of the Ge $\Gamma_{7c}$ and $L_{6c}$ CB edge states in Ge$_{1-x}$Sn$_{x}$. In a real alloy, all states that can (by symmetry) mix will mix: this physical effect will manifest in electronic calculations regardless of the choice of supercell. For example, Eckhardt et al.~\cite{Eckhardt_PRB_2014} noted the presence of strong Sn-induced $\Gamma_{7c}$-L$_{6c}$ mixing, but subsequently chose to analyse supercells in which the L points do not fold to the supercell zone centre (\textbf{K} = 0), to block this mixing and simplify their interpretation of the indirect- to direct-gap transition. However, unfolded band structures presented by Polak et al.~\cite{Polak_JPDAP_2017} for the same supercells revealed clear signatures of Sn-induced band mixing, occurring instead with states along the $\Lambda$ direction which did fold to \textbf{K} = 0. That is, the choice to specifically neglect $\Gamma_{7c}$-L$_{6c}$ mixing does not remove Sn-induced mixing of Ge states lying close in energy to the CB edge.

In the case of Ge$_{1-x}$Sn$_{x}$, we note that (i) the alloy CB edge originates from the Ge L$_{6c}$ states, and (ii) we are interested in the acquisition of direct Ge $\Gamma_{7c}$ character by the alloy CB edge. These observations suggest that there does not exist sufficient physical justification to explicitly neglect $\Gamma_{7c}$-L$_{6c}$ mixing -- which can be expected to be pronounced based on the small separation in energy of these states -- in the analysis of the Ge$_{1-x}$Sn$_{x}$ electronic structure. We therefore choose supercells in which the L points in the primitive unit cell Brillouin zone of the underlying diamond lattice fold to the supercell zone centre \textbf{K} = 0. This is the case in $n \times n \times n$ face-centred cubic (FCC) and simple cubic (SC) supercells for even values of $n$. As such, we employ $2 \times 2 \times 2$ FCC (16-atom) and SC (64-atom) supercells in our alloy electronic structure calculations.

% Table 2

\begin{table*}[t!]
    \caption{\label{tab:relaxation} Relaxed lattice constant $a$ and Ge-Sn nearest-neighbour bond length $r_{\protect\scalebox{0.7}{\textrm{Ge-Sn}}}$, for a series of ordered and disordered Ge$_{1-x}$Sn$_{x}$ alloy supercells relaxed via DFT using HSEsol and LDA XC functionals, and via the semi-empirical VFF potential of Eq.~\eqref{eq:vff_potential} (parametrised using HSEsol-calculated lattice and elastic constants for Ge, $\alpha$-Sn and zb-GeSn). The equilibrium lattice constant of Ge -- calculated using the HSEsol and LDA approaches, and taken as input to the VFF potential -- are provided for reference. For the disordered Ge$_{60}$Sn$_{4}$ supercell $r_{\protect\scalebox{0.7}{\textrm{Ge-Sn}}}$ refers to the average length of all relaxed Ge-Sn bonds; values listed in parentheses are the standard deviations associated with these averaged values.}
    \begin{ruledtabular}
        \begin{tabular}{ccc|ccc|ccc}
             & & & & $a$ (\AA) & & & $r_{\scalebox{0.7}{\textrm{Ge-Sn}}}$ (\AA) & \\
            Supercell          & $x$ (\%) & Description ~~~ & HSEsol & LDA    & VFF     ~~~~ & HSEsol & LDA    & VFF   \\
            \hline
            Ge                 & -----    & Pure Ge       ~~~ & 5.646  & 5.647  & 5.646 ~~~~ & -----  & -----  & ----- \\
            Ge$_{63}$ Sn$_{1}$ & 1.56     & Ordered SC    ~~~ & 5.661  & 5.664  & 5.658 ~~~~ & 2.574  & 2.578  & 2.558 \\
            Ge$_{15}$ Sn$_{1}$ & 6.25     & Ordered FCC   ~~~ & 5.702  & 5.703  & 5.694 ~~~~ & 2.554  & 2.554  & 2.544 \\
            Ge$_{60}$ Sn$_{4}$ & 6.25     & Disordered SC ~~~ & 5.701  & 5.700  & 5.694 ~~~~ & 2.583 (0.010)   & 2.584 (0.010) & 2.563 (0.010)
        \end{tabular}
    \end{ruledtabular}
\end{table*}

%%%%%%%%%%%%%%%%%
%%%% Results %%%%
%%%%%%%%%%%%%%%%%

\section{Results}
\label{sec:results}

Having outlined the theoretical models we have established to calculate the structural and electronic properties of Ge$_{1-x}$Sn$_{x}$ alloys, we now turn our attention to the application of these models in alloy supercell calculations. We compare and contrast the results of calculations carried out using all three models, reaffirming key features of the alloy properties and establishing the accuracy of the models. We begin in Sec.~\ref{sec:results_relaxation} by considering structural relaxation, before considering key features of the alloy supercell band structure in Sec.~\ref{sec:results_band_structure}.

%%%%%%%%%%%%%%%%%%%%%%%%%%%%%%%
%%%% Structural relaxation %%%%
%%%%%%%%%%%%%%%%%%%%%%%%%%%%%%%

\subsection{Structural relaxation}
\label{sec:results_relaxation}

\subsubsection{Alloy lattice constant}

Beginning with the lattice constants calculated using the HSEsol and LDA XC potentials for Ge, $\alpha$-Sn and zb-GeSn (cf.~Table~\ref{tab:dft_benchmark}), we note that these are in accordance with recent first principles calculations \cite{Eckhardt_PRB_2014} which suggested the presence of a small negative bowing parameter for the Ge$_{1-x}$Sn$_{x}$ alloy lattice constant. The presence of a negative bowing parameter implies that the alloy lattice constant at a given Sn composition $x$ is larger than that obtained by interpolating linearly between the lattice constants of Ge and $\alpha$-Sn. Indeed, the HSEsol-calculated zb-GeSn lattice constant $a = 6.079$ \AA~is close to, but in excess of, the average value 6.071 \AA~of those calculated for Ge and $\alpha$-Sn.

We do not pursude a detailed discussion of the evolution of the Ge$_{1-x}$Sn$_{x}$ structural properties with $x$ here, but instead focus on comparison of the lattice constants and atomic positions obtained via HSEsol, LDA and VFF relaxations of specific Ge$_{1-x}$Sn$_{x}$ alloy supercells. Specifically, we consider three distinct $2 \times 2 \times 2$ SC and FCC supercells: (i) an ordered 64-atom Ge$_{63}$Sn$_{1}$ SC supercell having $x = 1.56$\%, (ii) an ordered 16-atom Ge$_{15}$Sn$_{1}$ FCC supercell having $x = 6.25$\%, and (iii) a disordered 64-atom Ge$_{60}$Sn$_{4}$ SC supercell having $x = 6.25$\%. The four Sn atoms in the Ge$_{60}$Sn$_{4}$ were substituted at randomly selected sites on the lattice of the Ge$_{64}$ host matrix supercell. We note that this disordered supercell contains a Sn-Sn ``pair'' -- i.e.~two Sn atoms which are nearest neighbours. The ordered Ge$_{15}$Sn$_{1}$ and disordered Ge$_{60}$Sn$_{4}$ supercells have the same Sn composition, thereby allowing the importance of alloy disorder effects in determining the material properties to be inferred via comparison of the results of calculations carried out for each.

% Supercell relaxation

\subsubsection{Internal relaxation: ordered supercells}

The results of the HSEsol, LDA and VFF alloy supercell relaxations are summarised in Table~\ref{tab:relaxation}, where the relaxed supercell lattice constant $a$ is provided, in addition to the relaxed bond length $r_{\scalebox{0.7}{\textrm{Ge-Sn}}}$ between nearest-neighbour Ge and Sn atoms. The $A_{1}$-symmetric relaxation about the Sn lattice site in the ordered supercells produces four Ge-Sn nearest-neighbour bonds of equal length. Due to the presence of multiple Ge-Sn nearest-neighbour bonds and broken symmetry in the disordered Ge$_{60}$Sn$_{4}$ supercell, it is not possible to unambiguously assign a single value to $r_{\scalebox{0.7}{\textrm{Ge-Sn}}}$. We instead assign $r_{\scalebox{0.7}{\textrm{Ge-Sn}}}$ as the average of all of the relaxed Ge-Sn bond lengths in the supercell, and quantify deviations from this average using the standard deviation $\sigma ( r_{\scalebox{0.7}{\textrm{Ge-Sn}}} )$ of the relaxed Ge-Sn bond lengths (listed in parentheses in Table~\ref{tab:relaxation}). A more general comparison of the HSEsol, LDA and VFF relaxations for this disordered supercell is provided in Fig.~\ref{fig:relaxation_comparison} below.

% Ordered supercells - lattice constant

Beginning with the ordered Ge$_{63}$Sn$_{1}$ and Ge$_{15}$Sn$_{1}$ supercells, the HSEsol relaxation results in respective increases of 0.015 and 0.056 \AA~in the lattice constant compared to that of unstrained Ge (cf.~Table~\ref{tab:relaxation}). In both cases these relaxed lattice constants $a$ slightly exceed those calculated based on a linear interpolation of the HSEsol-calculated Ge and $\alpha$-Sn lattice constants, again suggesting the presence of a small, negative bowing parameter for $a$. \cite{Eckhardt_PRB_2014} Relaxing these supercells in the LDA yields respective increases of 0.017 and 0.056 \AA~in $a$ for the Ge$_{63}$Sn$_{1}$ and Ge$_{15}$Sn$_{1}$ supercells. While the calculated increase in $a$ for the Ge$_{15}$Sn$_{1}$ supercell precisely matches that obtained from the HSEsol calculation, the LDA relaxation overestimates the increase in $a$ due to Sn incorporation in the larger Ge$_{63}$Sn$_{1}$ supercell. This behaviour is somewhat surprising: the LDA-calculated zb-GeSn lattice constant is smaller than that obtained from HSEsol calculations, upon which basis it could be expected that nearest-neighbour Ge-Sn bonds relax to shorter lengths in LDA calculations. In spite of this, the LDA-relaxed lattice constants show very close agreement with the HSEsol values: the differences between the LDA and HSEsol relaxed lattice constants are, respectively, only 0.05 and 0.02\% for the Ge$_{63}$Sn$_{1}$ and Ge$_{15}$Sn$_{1}$ supercells.

Relaxing the same supercells using the VFF potential of Eq.~\eqref{eq:vff_potential} -- which, we recall, is parametrised based on HSEsol-calculated lattice and elastic constants -- the corresponding calculated increases in $a$ for Ge$_{63}$Sn$_{1}$ and Ge$_{15}$Sn$_{1}$ are 0.012 and 0.048 \AA. These values are in excellent agreement with the HSEsol relaxation: for Ge$_{63}$Sn$_{1}$ and Ge$_{15}$Sn$_{1}$ the respective differences between the VFF and HSEsol relaxed lattice constants are only 0.05 and 0.14~\%. Overall, for these ordered supercells we note that the LDA and VFF structural relaxations faithfully reproduce $a$ obtained from a HSEsol relaxation but, nevertheless, the LDA (VFF) relaxation tends to overestimate (underestimate) the increase in $a$ associated with substitutional Sn incorporation.

% Ordered supercells - bond lengths

Considering now the relaxed Ge-Sn nearest-neighbour bond lengths (cf.~Table~\ref{tab:relaxation}), we calculate $r_{\scalebox{0.7}{\textrm{Ge-Sn}}} = 2.574$ and 2.554 \AA~in HSEsol relaxations of the Ge$_{63}$Sn$_{1}$ and Ge$_{15}$Sn$_{1}$ supercells. Compared to the unstrained Ge nearest-neighbour bond length $r_{\scalebox{0.7}{\textrm{Ge}}}^{(0)} = 2.445$ \AA, these bond lengths represent respective increases of 0.129 and 0.109 \AA. This result is initially surprising: the relaxed lattice constants obtained from the HSEsol relaxation exhibit the opposite trend, with the increase in $a$ for the Ge$_{63}$Sn$_{1}$ supercell exceeding that calculated for Ge$_{15}$Sn$_{1}$. Further analysis reveals that this behaviour arises from the significantly smaller volume of the Ge$_{15}$Sn$_{1}$ supercell, in which Sn incorporation is accommodated by an increase in the average relaxed Ge-Ge bond length $r_{\scalebox{0.7}{\textrm{Ge-Ge}}}$. For the Ge$_{63}$Sn$_{1}$ and Ge$_{15}$Sn$_{1}$ supercells we respectively calculate $r_{\scalebox{0.7}{\textrm{Ge-Ge}}} = 2.448$ and 2.457 \AA. The LDA relaxations of these ordered supercells produce respective relaxed Ge-Sn bond lengths which are increased by 0.133 and 0.109 \AA~compared to the corresponding LDA-calculated equilibrium Ge bond length. The corresponding increases in the VFF relaxations of these supercells are 0.113 and 0.099 \AA. Using the HSEsol-calculated values as a reference, the computed differences in the LDA-relaxed values $r_{\scalebox{0.7}{\textrm{Ge-Sn}}}$ for the Ge$_{63}$Sn$_{1}$ and Ge$_{15}$Sn$_{1}$ supercells are, respectively, 0.15 and 0.01\%. The corresponding differences between the HSEsol and VFF relaxed Ge-Sn nearest-neighbour bond lengths for these supercells are 0.6 and 0.4\%.

 % Figure 2

 \begin{figure*}[t!]
 	\includegraphics[width=0.99\textwidth]{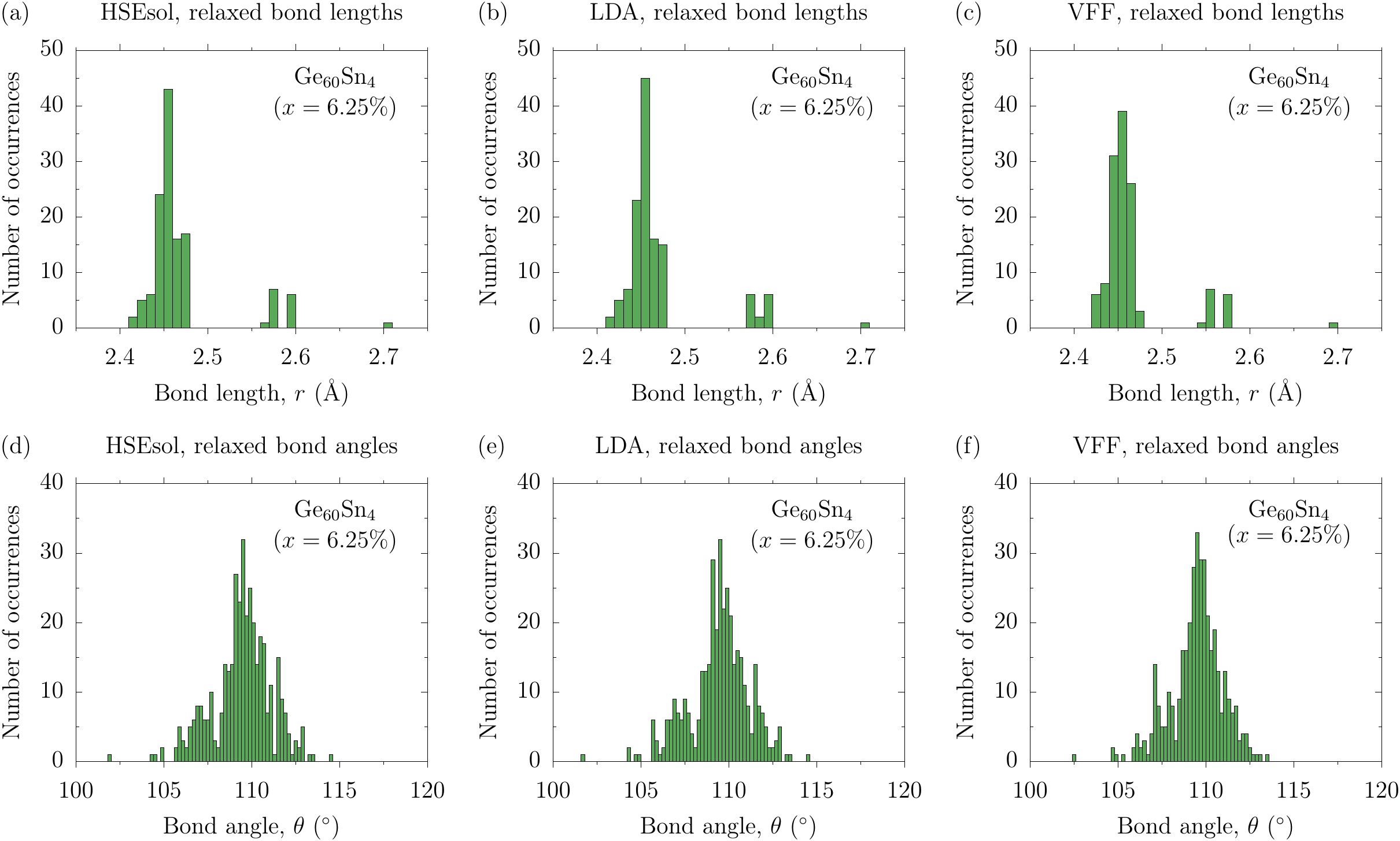}
 	\caption{Comparison of first principles and semi-empirical relaxations of the atomic positions for a disordered Ge$_{60}$Sn$_{4}$ ($x = 6.25$\%) supercell containing a Sn-Sn nearest-neighbour pair. Top row: relaxed nearest-neighbour bond lengths $r$ obtained from (a) HSEsol, (b) LDA, and (c) VFF minimisations of the lattice free energy. Bottom row: relaxed bond angles -- formed by adjacent pairs of nearest-neighbour bonds -- obtained from (d) HSEsol, (e) LDA, and (f) VFF minimisations of the lattice free energy. The relaxed bond lengths and angles have respectively been sorted into bins of width $\Delta r = 0.01$ \AA~and $\Delta \theta = 0.2^{\circ}$.}
 	\label{fig:relaxation_comparison}
 \end{figure*}

% Disordered supercell

\subsubsection{Internal relaxation: disordered supercell}

Turning to the disordered Ge$_{60}$Sn$_{4}$ supercell, we note that the relaxed lattice constants obtained from the HSEsol and LDA relaxations are slightly lower than those computed for the ordered Ge$_{15}$Sn$_{1}$ supercell having the same Sn composition. This slightly reduction in relaxed lattice constant in the presence of alloy disorder is likely related to the volume effect described above, whereby the larger surrounding Ge matrix in a 64-atom supercell provides more surrounding volume for expansion to accommodate the local strain introduced by Sn incorporation. The average relaxed Ge-Sn nearest-neighbour bond length $r_{\scalebox{0.7}{\textrm{Ge-Sn}}} = 2.583 \pm 0.010$ \AA~for this disordered supercell is significantly larger than that calculated for either of the ordered supercells. This is in part a result of the presence of a Sn-Sn nearest-neighbour pair in the supercell, the large local expansion of the lattice associated with which leads to larger relaxed Ge-Sn bond lengths about these two Sn atoms. Examining Table~\ref{tab:relaxation}, we note that this general trend is also observed in the LDA and VFF relaxations. However, the averaged nearest-neighbour bond lengths presented in Table~\ref{tab:relaxation} are insufficient in general to quantify the lattice relaxation in a disordered supercell. Thus, we present in Fig.~\ref{fig:relaxation_comparison} a more comprehensive view of the HSEsol, LDA and VFF relaxations of the Ge$_{60}$Sn$_{4}$ supercell. The top (bottom) row of Fig.~\ref{fig:relaxation_comparison} shows the relaxed bond lengths (angles) in the Ge$_{60}$Sn$_{4}$ supercell, sorted into bins of width $\Delta r = 0.01$ \AA~($\Delta \theta = 0.2^{\circ}$). Panels (a) and (d), (b) and (e), and (c) and (f) respectively show the distributions of relaxed bond lengths and angles obtained from the HSEsol, LDA and VFF relaxations.

Examining firstly the HSEsol results in Figs.~\ref{fig:relaxation_comparison}(a) and (d), we note the distribution of bond lengths brought about the the presence of alloy disorder. The three distinct peaks in the distribution of relaxed bond lengths in Fig.~\ref{fig:relaxation_comparison}(a) describe, in order of increasing bond length, relaxed Ge-Ge, Ge-Sn and Sn-Sn bonds. The width of the distribution of Ge-Ge bonds is strongly enhanced compared to that in the ordered Ge$_{63}$Sn$_{1}$ supercell: the computed standard deviation of the relaxed Ge-Ge bond lengths in the Ge$_{60}$Sn$_{4}$ supercell is 0.014 \AA, compared to 0.006 \AA~in the Ge$_{63}$Sn$_{1}$ supercell. As described above, the reduction in symmetry associated with the presence of alloy disorder also generates a distribution of relaxed Ge-Sn bond lengths, the standard deviation of which we compute to be 0.01 \AA. The single Sn-Sn bond in the supercell has a relaxed bond length $r_{\scalebox{0.7}{\textrm{Sn-Sn}}} = 2.704$ \AA, an increase of 0.259 \AA~compared to $r_{\scalebox{0.7}{\textrm{Ge}}}^{(0)}$. However, this relaxed Sn-Sn bond length is significantly lower than the equilibrium $\alpha$-Sn bond length $r_{\scalebox{0.7}{\textrm{Sn}}}^{(0)} = 2.813$ \AA. We attribute this compression of the Sn-Sn bond to the lower bulk modulus of $\alpha$-Sn compared to that of either Ge or zb-GeSn -- i.e.~the relatively soft Sn-Sn bond is compressed by the comparatively harder surrounding Ge-Ge and Ge-Sn bonds.

Figure~\ref{fig:relaxation_comparison}(d) shows the corresponding HSEsol-calculated distribution of relaxed bond angles in the Ge$_{60}$Sn$_{4}$ supercell. Since the equilibrium tetrahedral bond angle $\theta^{(0)} = 109.5^{\circ}$ is equal for all tetrahedrally-bonded constituent materials -- and hence the constituent materials Ge, $\alpha$-Sn and zb-GeSn -- the bond angle distribution provides, in general, less insight into the details of the lattice relaxation. We find the the relaxed bond angle distribution to be centred at $\theta^{(0)}$, with the lattice relaxation tending to produces a normal-like distribution of bond angles about this value. We note however the presence of a single outlying bond angle on the low side of the peak. This smallest angle describes a single Sn-Ge-Sn nearest-neighbour configuration, which has been highly distorted by the asymmetric relaxation of its local atomic environment.

Considering the corresponding LDA-calculated bond length distribution of Fig.~\ref{fig:relaxation_comparison}(b), we note excellent qualitative and quantitative agreement with the corresponding HSEsol distribution. Relative to the HSEsol-relaxed supercell we calculate a maximum difference of 0.14\% in the LDA-relaxed length of an individual nearest-neighbour bond, and an average error of only 0.04\% across all nearest-neighbour bonds in the supercell. However, evident in Fig.~\ref{fig:relaxation_comparison}(b) is a slight decrease in the width of the peak in the distribution of relaxed Ge-Sn bond lengths. We interpret this result as further confirmation that the LDA disproportionately overbinds Ge-Sn bonds compared to Ge-Ge or Sn-Sn bonds, as reflected in the differences between the HSEsol- and LDA-calculated lattice constants. Qualitatively, the VFF-calculated bond length distribution of Fig.~\ref{fig:relaxation_comparison}(c) is in excellent agreement with the HSEsol distribution of Fig.~\ref{fig:relaxation_comparison}(a): the VFF relaxation reproduces faithfully the width and relative number of relaxed Ge-Sn bonds having slightly different bond lengths. The maximum individual and overall average differences in relaxed bond length between the HSEsol and VFF relaxations are, respectively, 0.9 and 0.7\%. Finally, we note that the LDA- and VFF-calculated bond angle distributions of Figs.~\ref{fig:relaxation_comparison}(e) and (f) display no significant qualitative or quantitative deviations from the corresponding HSEsol-calculated distribution of Fig.~\ref{fig:relaxation_comparison}(d).

On the basis of these detailed comparisons, we conclude overall that both the LDA and VFF relaxations of a given Ge$_{1-x}$Sn$_{x}$ alloy supercell are in excellent quantitative agreement with the results of the significantly more computationally expensive HSEsol relaxation. This indicates that either LDA-DFT or the semi-empirical VFF potential of Eq.~\eqref{eq:vff_potential} can be used to reliably perform structural relaxations, offering high accuracy in conjunction with significantly reduced computational cost compared to HSEsol-DFT.

%%%%%%%%%%%%%%%%%%%%%%%%%%%%%%%%%%
%%%% Supercell band structure %%%%
%%%%%%%%%%%%%%%%%%%%%%%%%%%%%%%%%%

\subsection{Supercell band structure}
\label{sec:results_band_structure}

% Table 3

\begin{table*}[t!]
	\caption{\label{tab:energy_gaps} Band gap $E_{g}$ and VB spin-orbit splitting energy $\Delta_{\protect\scalebox{0.6}{\textrm{SO}}}$ for the Ge$_{1-x}$Sn$_{x}$ alloy supercells of Table~\ref{tab:relaxation}, calculated via DFT, using the HSEsol and mBJ XC functionals, and via a semi-empirical $sp^{3}s^{\ast}$ TB Hamiltonian. The corresponding calculated values of the direct $\Gamma_{8v}$-$\Gamma_{7c}$ and indirect (fundamental) $\Gamma_{8v}$-L$_{6c}$ band gaps of Ge are provided for reference, with the latter listed in parentheses.}
	\begin{ruledtabular}
		\begin{tabular}{ccc|ccc|ccc}
			 & & & & $E_{g}$ (eV) & & & $\Delta_{\scalebox{0.6}{\textrm{SO}}}$ (eV) & \\
			Supercell          & $x$ (\%) & Description   & HSEsol        & mBJ           & $sp^{3}s^{\ast}$ TB    & HSEsol & mBJ   & $sp^{3}s^{\ast}$ TB \\
			\hline
			Ge                 & -----    & Pure Ge       & 0.909 (0.766) & 0.868 (0.724) & 0.909 (0.766)       & 0.322     & 0.274 & 0.322               \\
			Ge$_{63}$ Sn$_{1}$ & 1.56     & Ordered SC    & 0.681         & 0.660         & 0.674               & 0.334     & 0.282 & 0.330               \\
			Ge$_{15}$ Sn$_{1}$ & 3.12     & Ordered FCC   & 0.388         & 0.356         & 0.392               & 0.379     & 0.316 & 0.353               \\
			Ge$_{60}$ Sn$_{4}$ & 6.25     & Disordered SC & 0.429         & 0.425         & 0.442               & 0.372     & 0.322 & 0.365
		\end{tabular}
	\end{ruledtabular}
\end{table*}

% Description of supercells

\subsubsection{Band gap and spin-orbit splitting energy}

Turning our attention to the electronic structure, we begin by considering the band structures calculated for the ordered Ge$_{63}$Sn$_{1}$ ($x = 1.56$\%) and Ge$_{15}$Sn$_{1}$ ($x = 6.25$\%) alloy supercells. The results of these calculations are shown in Fig.~\ref{fig:supercell_band_structure}, where the top (bottom) row shows the band structure calculated for Ge$_{63}$Sn$_{1}$ (Ge$_{15}$Sn$_{1}$) using, from left to right, the HSEsol ((a) and (d)), LDA + mBJ ((b) and (e)), and VFF + TB ((c) and (f)) models. In each band structure plot the left- and right-hand panels respectively show the supercell band dispersion calculated along the (111) and (001) directions in the supercell Brillouin zone. The supercell wave vector \textbf{K} is specified in units of $\frac{\pi}{A}$, where $A = 2a$ ($A = a$) is the lattice constant associated with the chosen $2 \times 2 \times 2$ SC (FCC) supercells. Recalling that the 64- and 16-atom supercells respectively possess SC and FCC lattice vectors, we note that the zone boundary along the (001) direction lies at $\frac{\pi}{A}$ in the 64-atom supercells. we recall that the supercells considered here have been chosen such so that L-points from the Brillouin zone of the primitive unit cell are folded to the supercell zone centre at $\textbf{K} = 0$. As such, the lowest energy CB states in Ge$_{64}$ and Ge$_{16}$ supercells are the degenerate L$_{6c}$ CB minima, folded to $\textbf{K} = 0$ from $\textbf{k} = \frac{\pi}{a} ( 1, 1, 1 )$ and equivalent points in the Brillouin zone of the primitive unit cell.

% Band gaps and spin-orbit splitting energies

Examining the results of the HSEsol band structure calculations for Ge$_{63}$Sn$_{1}$ and Ge$_{15}$Sn$_{1}$ in Figs.~\ref{fig:supercell_band_structure}(a) and~\ref{fig:supercell_band_structure}(d), we note firstly that Sn incorporation gives rise to a strong reduction of the supercell band gap $E_{g}$ with increasing $x$. For Ge$_{63}$Sn$_{1}$ and Ge$_{15}$Sn$_{1}$ we respectively calculate $E_{g} = 0.681$ and 0.388 eV, representing respective decreases of 85 and 378 meV compared to the indirect (fundamental) 0.766 eV $\Gamma_{8v}$-$\Gamma_{7c}$ band gap of Ge. We note also a moderate increase of the VB spin-orbit splitting energy $\Delta_{\scalebox{0.6}{\textrm{SO}}}$ with increasing $x$. The respective values $\Delta_{\scalebox{0.6}{\textrm{SO}}} = 0.334$ and 0.379 eV calculated for the ordered 64- and 16-atom alloy supercells are 12 and 57 meV larger than that calculated for Ge.

The values of $E_{g}$ and $\Delta_{\scalebox{0.6}{\textrm{SO}}}$ calculated using the HSEsol, LDA + mBJ and VFF + TB models are summarised in Table~\ref{tab:energy_gaps}. Considering the mBJ-calculated values, we note  slight underestimation of both the decrease in $E_{g}$ and increase in $\Delta_{\scalebox{0.6}{\textrm{SO}}}$ compared to HSEsol calculations. The mBJ-calculated Ge$_{63}$Sn$_{1}$ and Ge$_{15}$Sn$_{1}$ band gaps of 0.660 and 0.356 meV represent respective decreases of 64 and 368 meV compared to the indirect (fundamental) Ge band gap of 0.724 eV. The corresponding VB spin-orbit splitting energies $\Delta_{\scalebox{0.6}{\textrm{SO}}} = 0.282$ and 0.316 eV represent respective increases of 8 and 42 meV compared to the Ge spin-orbit splitting energy $\Delta_{\scalebox{0.6}{\textrm{SO}}} = 0.274$ eV. For the Ge$_{63}$Sn$_{1}$ supercell the VFF + TB model predicts a 92 meV decrease in $E_{g}$, which is slightly larger than the HSEsol- and mBJ-calculated 85 and 64 meV decrease. For the Ge$_{15}$Sn$_{1}$ supercell the TB-calculated band gap $E_{g} = 0.392$ eV is in excellent agreement with the HSEsol-calculated value $E_{g} = 0.388$ eV.

Considering the values of $E_{g}$ calculated using the three models for the disordered Ge$_{60}$Sn$_{4}$ supercell we note that, in all cases, $E_{g}$ is significantly larger than in the ordered Ge$_{15}$Sn$_{1}$ supercell having the same Sn composition ($x = 6.25$\%). This strong dependence of the calculated band gap on the alloy microstructure provides initial evidence that alloy disorder and related band mixing effects play an important role in determining the details of the alloy electronic structure. The corresponding differences between the values of $\Delta_{\scalebox{0.6}{\textrm{SO}}}$ calculated for the disordered Ge$_{60}$Sn$_{4}$ and ordered Ge$_{15}$Sn$_{1}$ supercells are significantly smaller, and further suggest that alloy disorder effects can be expected to primarily impact the CB structure in Ge$_{1-x}$Sn$_{x}$ alloys. Overall, we note that the trends in $E_{g}$ and $\Delta_{\scalebox{0.6}{\textrm{SO}}}$ vs.~$x$ calculated using all three models are in good quantitative agreement with previously published values. \cite{Eckhardt_PRB_2014,Polak_JPDAP_2017}

% Band structure

\subsubsection{Impact of Sn incorporation on conduction band structure}

\begin{figure*}[ht!]
	\includegraphics[width=0.99\textwidth]{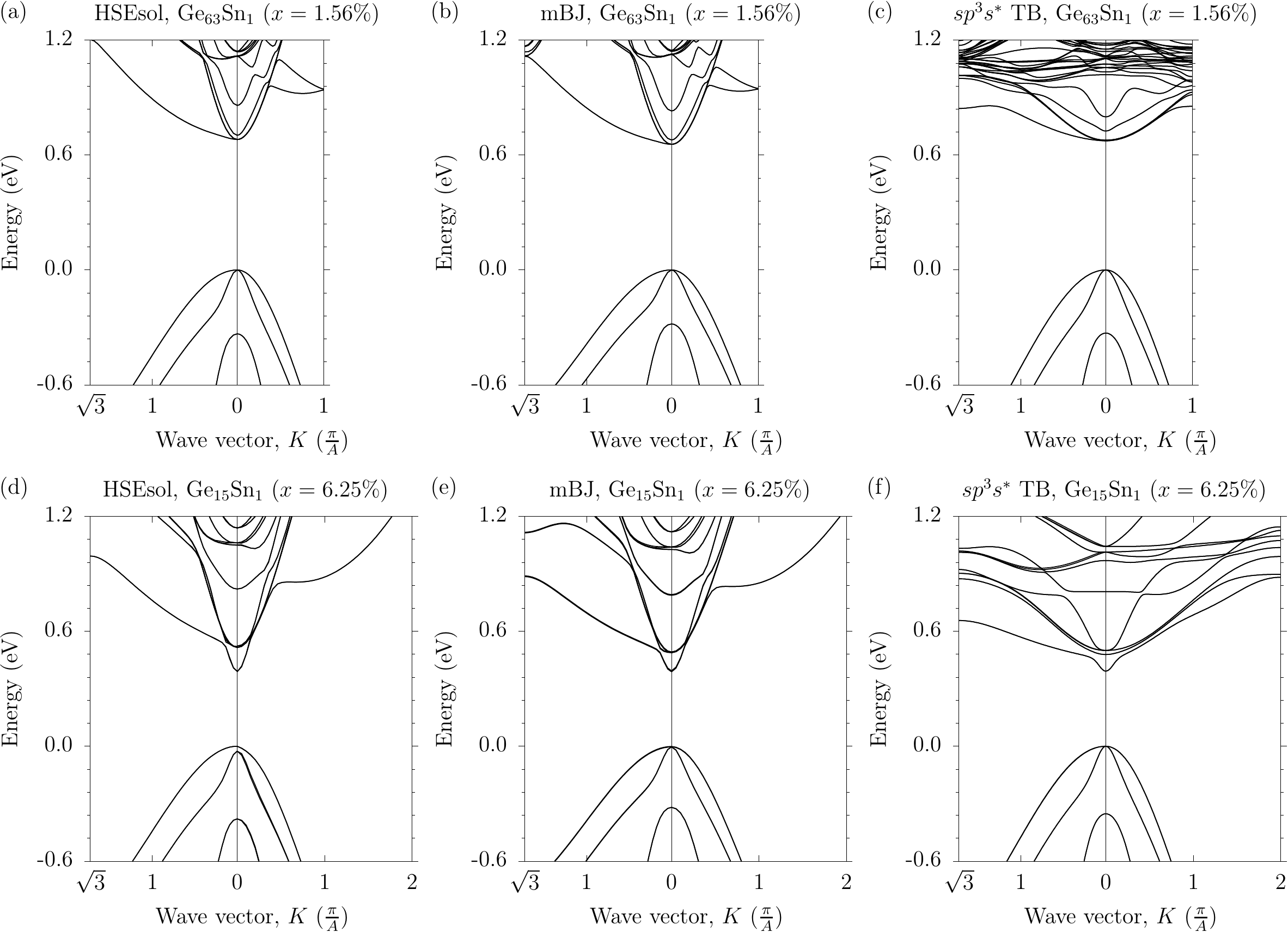}
	\caption{First principles and semi-empirical calculations of the band structure of two ordered Ge$_{1-x}$Sn$_{x}$ alloy supercells: a 64-atom $2 \times 2 \times 2$ SC Ge$_{63}$Sn$_{1}$ supercell (top row) and a 16-atom $2 \times 2 \times 2$ FCC Ge$_{15}$Sn$_{1}$ supercell (bottom row), having respective Sn compositions $x = 1.56$ and 6.25\%. The left- and right-hand panel of each plot respectively shows the band structure calculated along the (111) and (001) directions in the supercell Brillouin zone. The band structures calculated via DFT, using HSEsol ((a) and (d)) and mBJ ((b) and (e)) XC functionals, and via a semi-empirical $sp^{3}s^{\ast}$ TB Hamiltonian ((c) and (f)). The atomic positions used in the HSEsol, mBJ and $sp^{3}s^{\ast}$ TB band structure calculations were respectively obtained from HSEsol, LDA and VFF supercell relaxations. For comparative purposes, the zero of energy has been chosen to lie at the VB edge in all cases.}
	\label{fig:supercell_band_structure}
\end{figure*}

In addition to the respective strong decrease and moderate increase of $E_{g}$ and $\Delta_{\scalebox{0.6}{\textrm{SO}}}$ with increasing $x$, we note key qualitative changes in the nature of the alloy CB edge states compared to those in pure Ge. The lowest energy supercell CB states at the zone centre in the equivalent SC Ge$_{64}$ and FCC Ge$_{16}$ supercells are the eightfold degenerate L$_{6c}$ CB minima: the result of Kramers-degenerate bands folding back to $\textbf{K} = \textbf{0}$ from the four equivalent L points in the Brillouin zone of the primitive unit cell. The degeneracy of these folded L$_{6c}$ states is lifted in the ordered alloy supercell calculations of Fig.~\ref{fig:supercell_band_structure}. Examining the HSEsol-calculated CB eigenstates of the Ge$_{63}$Sn$_{1}$ supercell we find that the lowest energy CB eigenstates at $\textbf{K} = \textbf{0}$ are sixfold (threefold and Kramers) degenerate, and possess purely $T_{2}$ symmetry ($p$-like orbital character) at the Sn lattice site. The second lowest energy CB eigenstate lies 22 meV above the CB edge in energy, is twofold (Kramers) degenerate, and possesses $A_{1}$ symmetry ($s$-like orbital character) at the Sn lattice site. The next highest energy CB state in the HSEsol calculation is again twofold degenerate, lies 178 meV above the CB edge, and possesses possesses $A_{1}$ symmetry ($s$-like orbital character) at the Sn lattice site. The corresponding energy differences in the mBJ and TB calculations -- between the CB edge and the second- and third-lowest energy sets of $\textbf{K} = \textbf{0}$ alloy CB states -- are, respectively, 23 and 174 meV, and 54 and 120 meV. The discrepancies in the TB-calculated energies of the higher lying CB states is a consequence of the chosen TB fit to the HSEsol-calculated zb-GeSn band structure (cf.~Sec.~\ref{sec:models_semi_empirical}), during which the primary aim was to describe the alloy band gap and hybridised character of the alloy CB edge eigenstates.

% Figure 3

This splitting of the folded L$_{6c}$ states of Ge into distinct sets of states which are $T_{2}$- and $A_{1}$-symmetric ($p$- and $s$-like) at the Sn lattice site indicates Sn-induced mixing of the supercell zone-centre eigenstates of the Ge$_{64}$ host matrix semiconductor, driven by (i) $A_{1}$ symmetric lattice relaxation about the Sn lattice site, and (ii) interactions between the $(4s)^{2}$ valence orbitals of Ge and the $(5s)^{2}$ valence orbitals of $\alpha$-Sn. We note similar features in the HSEsol-calculated band structure of the Ge$_{15}$Sn$_{1}$ supercell (cf.~Fig.~\ref{fig:supercell_band_structure}). In this case the CB edge eigenstate is twofold (Kramers) degenerate and possesses purely $A_{1}$ symmetry ($s$-like orbital character) at the Sn lattice site, while the second lowest energy CB eigenstates lie 127 meV above the CB edge, are sixfold (threefold and Kramers) degenerate and possess purely $T_{2}$ symmetry ($p$-like orbital character) at the Sn lattice site.  The corresponding energy differences in the mBJ and TB calculations are, respectively, 100 and 104 meV. We note that the $T_{2}$-symmetric ($p$-like) triplet in the HSEsol and mBJ calculations splits into a singlet and higher lying doublet -- separated by 21 meV -- in the TB calculation. This is a result of the large spin-orbit coupling associated with the substitutional Sn atom, combined with the manner in which the atomic spin-orbit coupling is parametrised in the TB method. \cite{Chadi_PRB_1977} This ordering of the lowest and second-lowest energy CB states is reversed compared to that in the Ge$_{63}$Sn$_{1}$ supercell. Further analysis of the CB edge eigenstates, presented below, shows that the direct nature of the Ge$_{1-x}$Sn$_{x}$ alloy CB edge states evolves continously with increasing $x$, and that this evolution is driven by the aforementioned Sn-induced mixing. This band mixing, which acts to hybridise the L$_{6c}$ and $\Gamma_{7c}$ CB edge states of Ge, leads to the emergence of a strongly hybridised alloy CB edge in Ge$_{1-x}$Sn$_{x}$. \cite{Eckhardt_PRB_2014} By comparison, the impact of Sn incorporation on the VB is minimal: the alloy VB edge states in all supercells retain primarily Ge $\Gamma_{8v}$ character.

While the TB model describes well the impact of Sn incorporation on the CB and VB edge states at $\textbf{K} = \textbf{0}$ compared to the DFT calculations we note, in both the 64- and 16-atom calculations of Figs.~\ref{fig:supercell_band_structure}(c) and~\ref{fig:supercell_band_structure}(f), (i) overestimation of electron effective masses, and (ii) the presence in the CB dispersion of additional folded bands within the range of energies displayed in Fig.~\ref{fig:supercell_band_structure}. These features represent a typical TB fit to the band structure. Firstly, fitting the TB parameters to the energies of the lowest CB at the L, $\Gamma$ and X points is known to lead to overestimation of CB edge effective masses in a nearest-neighbour $sp^{3}s^{\ast}$ model. \cite{Boykin_PRB_1997} Secondly, the TB structure described by the $sp^{3}s^{\ast}$ model is dispersionless between the X and W points in the Brillouin zone: \cite{Vogl_JPCS_1983,Chadi_PSSB_1975} energies of W-point CB states generally lie at higher energies than those at X in first principles calculations, leading to more bands folding back to the supercell zone centre at lower energies in the TB calculations. We emphasise however that these well-known features of the $sp^{3}s^{\ast}$ TB model do not impede the description of the character, hybridisation and localisation of zone-centre alloy band edge states. The TB method allows for accurate descriptions of the impact of localised impurities, as we have demonstrated previously for highly-mismatched III-V-N \cite{Lindsay_SSC_1999,Reilly_SST_2002} and III-V-Bi \cite{Usman_PRB_2011,Usman_PRA_2018} alloys, as well as III-N semiconductor alloys. \cite{Schulz_APE_2013,Caro_PRB_2013} We emphasise that both the mBJ and TB calculations qualitatively and quantitatively describe the behaviour revealed by the HSEsol calculations, providing a key confirmation of their suitability to perform accurate analysis of the Ge$_{1-x}$Sn$_{x}$ electronic structure.

% Pressure coefficients

\subsubsection{Sn-induced band mixing: band gap pressure coefficient}

To quantify the Sn-induced band mixing and its impact on the nature of the alloy CB edge states, we have used all three models to calculate the pressure coefficient $\frac{dE_{g}}{dP}$ of the fundamental band gap. To do so we proceed by applying hydrostatic pressure to the supercell and relaxing the internal degrees of freedom before computing the electronic structure. The results of these calculations are summarised in Table~\ref{tab:pressure_coefficients}. HSEsol calculations predict $\frac{dE_{g}}{dP} = 4.66$ and 13.33 meV kbar$^{-1}$ for the indirect $\Gamma_{8v}$-L$_{6c}$ and direct $\Gamma_{8v}$-$\Gamma_{7c}$ band gaps of Ge. These calculated values are in good quantitative agreement with the respective measured values of 4.3 and 12.9 meV kbar$^{-1}$. \cite{Eales_PW_2017,Eales_submitted_2019} The corresponding LDA + mBJ-calculated values 4.07 and 13.23 meV kbar$^{-1}$ are in good quantitative agreement with both the HSEsol calculations and experiment. Using the VFF + TB model we calculate indirect- and direct-gap pressure coefficients of 4.69 and 13.52 meV kbar$^{-1}$ for Ge. The TB parameters are fitted exactly to the HSEsol band edge hydrostatic deformation potentials. We note the small ($< 1.5$\%) errors in the TB-calculated pressure coefficients reflects the use of a harmonic VFF potential to relax the internal degrees of freedom under applied pressure. While the VFF potential of Eq.~\eqref{eq:vff_potential} describes exactly the bulk moduli $B$ of Ge, $\alpha$-Sn and zb-GeSn, it does not describe the pressure dependence of $B$ -- which are captured implicitly in the DFT calculations -- leading to minor errors at experimentally relevant pressures.

Given the large differences in the pressure coefficients associated with the direct and indirect ($\Gamma_{8v}$-L$_{6c}$ and $\Gamma_{8v}$-X$_{5c}$) band gaps of Ge, calculations and experimental measurements of $\frac{dE_{g}}{dP}$ constitute a key experimental signature of band mixing effects: hybridised CB edge states possess a pressure coefficient intermediate between those of the band gaps associated with the constituent (hybridising) states. HSEsol calculations predict respective pressure coefficients $\frac{dE_{g}}{dP} = 4.75$ and 10.00 meV kbar$^{-1}$ for the ordered Ge$_{63}$Sn$_{1}$ and Ge$_{15}$Sn$_{1}$ supercells. The first of these values is very close to the indirect-gap pressure coefficient of Ge, reflecting that the CB edge states in Ge$_{63}$Sn$_{1}$ are largely derived from the L$_{6c}$ CB edge states of the corresponding Ge$_{64}$ host matrix semiconductor. The second of these values is intermediate between those calculated for the indirect (fundamental) $\Gamma_{8v}$-L$_{6c}$ and direct band gaps of Ge, confirming the presence of Sn-induced $\Gamma_{7c}$-L$_{6c}$ mixing. That is, as $x$ increases to 6.25\% the calculated pressure coefficient increases significantly, towards that of the Ge direct band gap. We note that the calculated values of $\frac{dE_{g}}{dP}$ for the Ge$_{15}$Sn$_{1}$ supercell are in good agreement with the measured value of 9.2 meV kbar$^{-1}$ for a Ge$_{0.94}$Sn$_{0.06}$ ($x = 6$\%) photodiode. \cite{Eales_PW_2017,Eales_submitted_2019} These calculated and measured values of $\frac{dE_{g}}{dP}$ suggest that the hybridised nature of the alloy CB edge states in Ge$_{1-x}$Sn$_{x}$ depends strongly on Sn composition $x$, and suggests that the CB edge states in the ordered Ge$_{15}$Sn$_{1}$ supercell are primarily Ge $\Gamma_{7c}$-derived, with the alloy having primarily (but not purely) direct-gap character.

% Table 4

\begin{table}[t!]
	\caption{\label{tab:pressure_coefficients} Band gap pressure coefficients for the Ge$_{1-x}$Sn$_{x}$ alloy supercells of Tables~\ref{tab:relaxation} and~\ref{tab:energy_gaps}, calculated using the HSEsol, LDA + mBJ and VFF + TB models described in Sec.~\ref{sec:theoretical_models}. The calculated pressure coefficients of the direct $\Gamma_{8v}$-$\Gamma_{7c}$ and indirect (fundamental) $\Gamma_{8v}$-L$_{6c}$ band gaps of Ge are provided for reference, with the latter listed in parentheses.}
	\begin{ruledtabular}
		\begin{tabular}{cc|ccc}
			& & & $\frac{dE_{g}}{dP}$ (meV kbar$^{-1}$) & \\
			Supercell          & $x$ (\%) & HSEsol          & mBJ            & $sp^{3}s^{\ast}$ TB \\
			\hline
			Ge                 & -----    &  $13.33$ (4.66) & $13.23$ (4.07) & $13.52$ (4.69)      \\
			Ge$_{63}$ Sn$_{1}$ & 1.56     &  $4.75$         & $4.19$         & $4.73$              \\
			Ge$_{15}$ Sn$_{1}$ & 6.25     & $10.00$         & $9.50$         & $10.00$             \\
			Ge$_{60}$ Sn$_{4}$ & 6.25     &  $8.32$         & $7.57$         & $9.39$              \\
		\end{tabular}
	\end{ruledtabular}
\end{table}

For the disordered Ge$_{60}$Sn$_{4}$ supercell we calculate $\frac{dE_{g}}{dP} = 8.32$ meV kbar$^{-1}$. This value is reduced compared to that calculated for the ordered Ge$_{15}$Sn$_{1}$ supercell having the same Sn composition, further emphasising that alloy disorder has a strong impact on the nature of the hybridised Ge$_{1-x}$Sn$_{x}$ CB edge states. This calculated variation of $\frac{dE_{g}}{dP}$ at fixed $x$ highlights that important quantitative differences can result in the character of the Ge$_{1-x}$Sn$_{x}$ CB edge states, resulting from an interplay of band mixing and alloy disorder effects. Examining the data of Table~\ref{tab:pressure_coefficients} we note that the LDA + mBJ and VFF + TB models correctly capture the increase in $\frac{dE_{g}}{dP}$ vs.~$x$ obtained from HSEsol calculations, suggesting that both models accurately capture the nature of the hybridised CB edge states and their evolution with increasing $x$.

Based on this analysis we reach three conclusions. Firstly, the calculated (and measured) values of $\frac{dE_{g}}{dP}$ being intermediate between those of the direct and $\Gamma_{8v}$-L$_{6c}$ band gaps of Ge indicates that the Ge$_{1-x}$Sn$_{x}$ alloy CB edge states consist of a strong admixture of Ge $\Gamma_{7c}$ and L$_{6c}$ states. Secondly, this band mixing is driven by the differences in covalent radius and chemical properties between Ge and Sn. Thirdly, this band mixing evolves continuously with increasing $x$, transferring Ge $\Gamma_{7c}$ character to the alloy band edge. This final conclusion is in accordance with the measurements of Refs.~\onlinecite{Eales_PW_2017} and~\onlinecite{Eales_submitted_2019}, which found $\frac{dE_{g}}{dP} = 4.33$, 9.2, 10.4 and 12.9 meV kbar$^{-1}$ for the fundamental band gap in Ge$_{1-x}$Sn$_{x}$ photodiodes having $x = 0$, 6, 8 and 10\%.

Contrary to the prevailing assumption in the literature, this suggests that the indirect- to direct-gap transition in Ge$_{1-x}$Sn$_{x}$ proceeds continuously rather than occurring abruptly at some critical Sn composition. Combined, these results suggest significant implications for the interpretation of the Ge$_{1-x}$Sn$_{x}$ band structure. The strongly hybridised nature of the Ge$_{1-x}$Sn$_{x}$ CB edge states can be expected to have significant consequences for technologically relevant material properties. The LDA + mBJ and VFF + TB models established and benchmarked here provide quantitatively accurate descriptions of band mixing effects in alloy supercell calculations.

%%%%%%%%%%%%%%%%%%%%%
%%%% Conclusions %%%%
%%%%%%%%%%%%%%%%%%%%%

\section{Conclusions}
\label{sec:conclusions}

In summary, we have established three atomistic models of the structural and electronic properties of Ge$_{1-x}$Sn$_{x}$ alloys. The first model uses HSEsol DFT to perform structural relaxation and to calculate the electronic structure. This provides a highly accurate description of (i) the constituent materials Ge, $\alpha$-Sn and zb-GeSn, and (ii) the hybridised nature of the Ge$_{1-x}$Sn$_{x}$ alloy CB edge states. The second model uses LDA structural relaxation combined with mBJ electronic structure calculations. The third, semi-empirical model uses VFF structural relaxation combined with $sp^{3}s^{\ast}$ TB electronic structure calculations. Since the HSEsol DFT calculations are in good quantitative agreement with experimental measurements, the results of these calculations were taken as benchmark for the other two models.

The trends in the Ge$_{1-x}$Sn$_{x}$ alloy lattice constant, relaxed atomic positions, band gap and VB spin-orbit splitting energy calculated using the LDA + mBJ and VFF + TB models were found to be in excellent quantitative agreement with the results of full HSEsol calculations. Alloy supercell band structure calculations carried out using all three models highlight the potential importance of Sn-induced band mixing and alloy disorder in determining the hybridised nature of the Ge$_{1-x}$Sn$_{x}$ CB structure. As it offers an measureable means by which to quantify this hybridisation, the band gap pressure coefficient was also calculated using the three approaches. Again, the LDA + mBJ and VFF + TB calculations were found to capture both qualitatively and quantitatively the trends observed in full HSEsol calculations. Our calculations support the recent suggestion that the evolution of a direct band gap in Ge$_{1-x}$Sn$_{x}$ is driven by Sn-induced band mixing and occurs continously with increasing $x$, rather than occuring abruptly at a single critical Sn composition. Further work is required to clarify the nature and evolution of the Ge$_{1-x}$Sn$_{x}$ electronic structure, and to identify associated consequences for material properties relevant to device applications. The models established here provide a suitable platform to perform such investigations.

Based on our benchmark calculations we reach three conclusions. Firstly, the LDA + mBJ model offers an accurate description of Ge$_{1-x}$Sn$_{x}$ alloys at reduced computational expense compared to HSEsol calculations, providing access to larger systems within a first principles framework. Secondly, the VFF + TB model provides a sufficiently accurate description of the alloy band structure close in energy to the CB and VB edges to justify its use to underpin future analyses, where it will provide access to systems several orders of magnitude larger than those that can currently be treated via DFT. Thirdly, the VFF potential of Eq.~\eqref{eq:vff_potential} -- parametrised analytically on the basis of HSEsol calculations -- provides a highly efficient means by which to perform structural relaxations for Ge$_{1-x}$Sn$_{x}$ alloy supercells, and is of sufficient accuracy that it can be reliably used to circumvent the requirement to perform computationally expensive DFT structural relaxation.

Overall, we conclude that electronic structure calculations for Ge$_{1-x}$Sn$_{x}$ alloys must explicitly include atomistic alloying effects -- which are omitted in widely-used VCA-based models -- to accurately account for the impact of Sn incorporation on key material parameters (e.g.~optical transition strengths, carrier mobilities and band-to-band tunneling rates). Given the expected importance of these effects in determining technologically relevant material properties, the establishment and benchmarking of appropriate atomistic theoretical models represents an important step to enable predictive theoretical analysis of proposed Ge$_{1-x}$Sn$_{x}$-based photonic, electronic and photovoltaic devices. The theoretical models we have presented allow for the treatment -- with miniminal loss of accuracy -- of larger systems than those accessible to DFT employing hybrid XC functionals, thereby providing a basis for direct atomistic calculations of the electronic, optical and transport properties of disordered Ge$_{1-x}$Sn$_{x}$ alloys and realistically-sized nanostructures.

%%%%%%%%%%%%%%%%%%%%%%%%%%
%%%% Acknowledgements %%%%
%%%%%%%%%%%%%%%%%%%%%%%%%%

\section*{Acknowledgements}

E.J.O'H.~and C.A.B.~contributed equally to this work. This work was supported by Science Foundation Ireland (SFI; project nos.~15/IA/3082, 14/IA/2513 and 13/SIRG/2210), and by the National University of Ireland (NUI; via the Post-Doctoral Fellowship in the Sciences, held by C.A.B.). The authors acknowledge the provision of computing resources by SFI via Tyndall National Institute and the Irish Centre for High-End Computing (ICHEC; additional support for which is provided by the Higher Education Authority, as well as the Departments of Education and Skills, and Business, Enterprise and Innovation of the Government of Ireland). The authors thank Dr.~Timothy D.~Eales, Dr.~Igor P.~Marko, and Prof.~Stephen J.~Sweeney (University of Surrey, U.K.) for providing access to the results of their experimental measurements prior to publication.

%%%%%%%%%%%%%%%%%%%%%%%%%%%%%%%
%%%% Data access statement %%%%
%%%%%%%%%%%%%%%%%%%%%%%%%%%%%%%

% \section*{Data access}

% The data associated with this work are openly available, and can be accessed via Ref.~\textcolor{red}{ZZZ}.

%%%%%%%%%%%%%%%%%%%%
%%%% References %%%%
%%%%%%%%%%%%%%%%%%%%

% \bibliographystyle{apsrev}      % Using the Physical Review style for references
% \bibliography     {GeSn_models} % The bibliography is contained in the file: GeSn_models.bib

\end{document}